\documentclass[12pt]{article}%
\usepackage{amsmath}
\usepackage{graphicx}
\usepackage{amsfonts}
\usepackage{amssymb}%
\newtheorem{theorem}{Theorem}
\newtheorem{acknowledgement}[theorem]{Acknowledgement}

\newtheorem{proposition}[theorem]{Proposition}

\newenvironment{proof}[1][Proof]{\textbf{#1.} }{\ \rule{0.5em}{0.5em}}
\begin{document}

\title{On a quaternionic Maxwell equation for the time-dependent electromagnetic
field in a chiral medium}
\author{Sergei M. Grudsky*, Kira V. Khmelnytskaya**, 
\and Vladislav V. Kravchenko**\\\ \ \ \ \ *Dept. of Mathematics\\CINVESTAV \\National Polytechnic Institute\\Mexico City\\\ \ \ \ \ \ \ **Dept. of Telecommunications,\\SEPI ESIME Zacatenco\\National Polytechnic Institute\\Av. IPN S/N, C.P.07738 D.F.\\Mexico}
\maketitle

\begin{abstract}
Maxwell's equations for the time-dependent electromagnetic field in a
homogeneous chiral medium are reduced to a single quaternionic equation. Its
fundamental solution satisfying the causality principle is obtained which
allows us to solve the time-dependent chiral Maxwell system with sources.

\end{abstract}

\section{Introduction}

We consider Maxwell's equations for the time-dependent electromagnetic field
in a homogeneous chiral medium and show their equivalence to a single
quaternionic equation. This result generalizes the well known (see \cite{Shn},
\cite{Imaeda}, \cite{AQA}) quaternionic reformulation of the Maxwell equations
for non-chiral media. Nevertheless the new quaternionic differential operator
is essentially different from the quaternionic operator corresponding to the
non-chiral case. We obtain a fundamental solution of the new operator in
explicit form satisfying the causality principle. Its convolution with a
quaternionic function representing sources of the electromagnetic field gives
us a solution of the inhomogeneous Maxwell system in a whole space.

\section{Maxwell's equations for chiral media\label{Maxw}}

Consider time-dependent Maxwell's equations%

\begin{equation}
\operatorname*{rot}\overrightarrow{E}(t,x)=-\partial_{t}\overrightarrow
{B}(t,x), \label{rot1}%
\end{equation}

\begin{equation}
\operatorname*{rot}\overrightarrow{H}(t,x)=\partial_{t}\overrightarrow
{D}(t,x)+\overrightarrow{j}(t,x), \label{rot2}%
\end{equation}

\begin{equation}
\operatorname*{div}\overrightarrow{E}(t,x)=\frac{\rho(t,x)}{\varepsilon
},\qquad\operatorname*{div}\overrightarrow{H}(t,x)=0 \label{div}%
\end{equation}
with the Drude-Born-Fedorov constitutive relations corresponding to the chiral
media \cite{ARS}, \cite{Lak}, \cite{LSihvola}
\begin{equation}
\overrightarrow{B}(t,x)=\mu(\overrightarrow{H}(t,x)+\beta\operatorname*{rot}%
\overrightarrow{H}(t,x)), \label{DBF1}%
\end{equation}%
\begin{equation}
\overrightarrow{D}(t,x)=\varepsilon(\overrightarrow{E}(t,x)+\beta
\operatorname*{rot}\overrightarrow{E}(t,x)), \label{DBF2}%
\end{equation}
where $\beta$ is the chirality measure of the medium. $\beta,\varepsilon,\mu$
are real scalars assumed to be constants. Note that the charge density $\rho$
and the current density $\overrightarrow{j}$ are related by the continuity
equation $\partial_{t}\rho+\operatorname*{div}\overrightarrow{j}=0$.

Incorporating the constitutive relations (\ref{DBF1}), (\ref{DBF2}) into the
system (\ref{rot1})-(\ref{div}) we arrive at the main object of our study, the
time-dependent Maxwell system for a homogeneous chiral medium%

\begin{equation}
\operatorname*{rot}\overrightarrow{H}(t,x)=\varepsilon(\partial_{t}%
\overrightarrow{E}(t,x)+\beta\partial_{t}\operatorname*{rot}\overrightarrow
{E}(t,x))+\overrightarrow{j}(t,x), \label{Max1}%
\end{equation}%
\begin{equation}
\operatorname*{rot}\overrightarrow{E}(t,x)=-\mu(\partial_{t}\overrightarrow
{H}(t,x)+\beta\partial_{t}\operatorname*{rot}\overrightarrow{H}(t,x)),
\label{Max2}%
\end{equation}

\begin{equation}
\operatorname*{div}\overrightarrow{E}(t,x)=\frac{\rho(t,x)}{\varepsilon
},\qquad\operatorname*{div}\overrightarrow{H}(t,x)=0. \label{Max3}%
\end{equation}

Application of $\operatorname*{rot}$ to (\ref{Max1}) and (\ref{Max2}) allows
us to separate the equations for $\overrightarrow{E}$ and $\overrightarrow{H}$
and to obtain in this way the wave equations for a chiral medium%
\begin{equation}
\operatorname*{rot}\operatorname*{rot}\overrightarrow{E}+\varepsilon
\mu\partial_{t}^{2}\overrightarrow{E}+2\beta\varepsilon\mu\partial_{t}%
^{2}\operatorname*{rot}\overrightarrow{E}+\beta^{2}\varepsilon\mu\partial
_{t}^{2}\operatorname*{rot}\operatorname*{rot}\overrightarrow{E}=-\mu
\partial_{t}\overrightarrow{j}-\beta\mu\partial_{t}\operatorname*{rot}%
\overrightarrow{j}, \label{wave1}%
\end{equation}%
\begin{equation}
\operatorname*{rot}\operatorname*{rot}\overrightarrow{H}+\varepsilon
\mu\partial_{t}^{2}\overrightarrow{H}+2\beta\varepsilon\mu\partial_{t}%
^{2}\operatorname*{rot}\overrightarrow{H}+\beta^{2}\varepsilon\mu\partial
_{t}^{2}\operatorname*{rot}\operatorname*{rot}\overrightarrow{H}%
=\operatorname*{rot}\overrightarrow{j}. \label{wave2}%
\end{equation}

It should be noted that when $\beta=0$, (\ref{wave1}) and (\ref{wave2}) reduce
to the wave equations for non-chiral media but in general to the difference of
the usual non-chiral wave equations their chiral generalizations represent
equations of fourth order.

\section{Some notations from quaternionic analysis}

We will consider biquaternion-valued functions defined in some domain
$\Omega\subset\mathbb{R}^{3}.$ On the set of continuously differentiable such
functions the well known Moisil-Teodoresco operator is defined by the
expression $D=i_{1}\frac{\partial}{\partial x_{1}}+i_{2}\frac{\partial
}{\partial x_{2}}+i_{3}\frac{\partial}{\partial x_{3}}$ (see, e.g.,
\cite{GS2}), where $i_{k}$, $k=1,2,3$ are basic quaternionic imaginary units.
Denote $D_{\alpha}=D+\alpha$, where $\alpha\in\mathbb{C}$ and
$\operatorname*{Im}\alpha\geq0$. The fundamental solution for this operator is
known \cite{Krdep} (see also \cite{AQA}):%

\begin{equation}
\mathcal{K}_{\alpha}(x)=-\operatorname*{grad}\Theta_{\alpha}(x)+\alpha
\Theta_{\alpha}(x)=(\alpha+\frac{x}{\left|  x\right|  ^{2}}-i\alpha\frac
{x}{\left|  x\right|  })\Theta_{\alpha}(x), \label{fund_d}%
\end{equation}
where $i$ is the usual complex imaginary unit commuting with $i_{k}$,
$x=\sum_{k=1}^{3}x_{k}i_{k}$ and $\Theta_{\alpha}(x)=-\frac{e^{i\alpha\left|
x\right|  }}{4\pi\left|  x\right|  }$. Note that $\mathcal{K}_{\alpha}$
fulfills the following radiation condition at infinity uniformly in all
directions%
\begin{equation}
(1+\frac{ix}{\left|  x\right|  })\cdot\mathcal{K}_{\alpha}(x)=o(\frac
{1}{\left|  x\right|  }),\qquad\text{when }\left|  x\right|  \rightarrow
\infty\label{rad}%
\end{equation}
which is in agreement with the Silver-M\"{u}ller radiation conditions
\cite{KKR1}.

\section{Field equations in quaternionic form}

In this section we rewrite the field equations from Section \ref{Maxw} in
quaternionic form.

Let us introduce the following quaternionic operator
\begin{equation}
M=\beta\sqrt{\varepsilon\mu}\partial_{t}D+\sqrt{\varepsilon\mu}\partial_{t}-iD
\label{A}%
\end{equation}
and consider the purely vectorial biquaternionic function
\begin{equation}
\overrightarrow{V}(t,x)=\overrightarrow{E}(t,x)-i\sqrt{\frac{\mu}{\varepsilon
}}\overrightarrow{H}(t,x). \label{V}%
\end{equation}

\begin{proposition}
The quaternionic equation
\begin{equation}
M\overrightarrow{V}(t,x)=-\sqrt{\frac{\mu}{\varepsilon}}\overrightarrow
{j}(t,x)-\beta\sqrt{\frac{\mu}{\varepsilon}}\partial_{t}\rho(t,x)+\frac
{i\rho(t,x)}{\varepsilon} \label{Amax}%
\end{equation}
is equivalent to the Maxwell system (\ref{Max1})-(\ref{Max3}), the vectors
$\overrightarrow{E}$ and $\overrightarrow{H}$ are solutions of (\ref{Max1}%
)-(\ref{Max3}) if and only if the purely vectorial biquaternionic function
$\overrightarrow{V}$ defined by (\ref{V}) is a solution of (\ref{Amax}).
\end{proposition}

\begin{proof}
The scalar and the vector parts of (\ref{Amax}) have the form%
\begin{equation}
-\beta\sqrt{\varepsilon\mu}\partial_{t}\operatorname*{div}\overrightarrow
{E}+\sqrt{\frac{\mu}{\varepsilon}}\operatorname*{div}\overrightarrow
{H}+i(\operatorname*{div}\overrightarrow{E}+\beta\mu\partial_{t}%
\operatorname*{div}\overrightarrow{H})=-\beta\sqrt{\frac{\mu}{\varepsilon}%
}\partial_{t}\rho+\frac{i\rho}{\varepsilon}, \label{sc}%
\end{equation}%
\begin{equation}
\beta\sqrt{\varepsilon\mu}\partial_{t}\operatorname*{rot}\overrightarrow
{E}+\sqrt{\varepsilon\mu}\partial_{t}\overrightarrow{E}-\sqrt{\frac{\mu
}{\varepsilon}}\operatorname*{rot}\overrightarrow{H}-i(\operatorname*{rot}%
\overrightarrow{E}+\beta\mu\partial_{t}\operatorname*{rot}\overrightarrow
{H}+\mu\partial_{t}\overrightarrow{H})=-\sqrt{\frac{\mu}{\varepsilon}%
}\overrightarrow{j}. \label{vec}%
\end{equation}
The real part of (\ref{vec}) coincides with (\ref{Max1}) and the imaginary
part coincides with (\ref{Max2}). Applying divergence to the equation
(\ref{vec}) and using the continuity equation gives us
\[
\partial_{t}\operatorname*{div}\overrightarrow{H}=0\text{\quad and\quad
}\partial_{t}\operatorname*{div}\overrightarrow{E}=\frac{1}{\varepsilon
}\partial_{t}\rho.
\]
Taking into account these two equalities we obtain from (\ref{sc}) that the
vectors $\overrightarrow{E}$ and $\overrightarrow{H}$ satisfy equations
(\ref{Max3}).
\end{proof}

It should be noted that for $\beta=0$ from (\ref{A}) we obtain the operator
which was studied in \cite{KKR} with the aid of the factorization of the wave
operator for non-chiral media%
\[
\varepsilon\mu\partial_{t}^{2}-\Delta_{x}=(\sqrt{\varepsilon\mu}\partial
_{t}+iD)(\sqrt{\varepsilon\mu}\partial_{t}-iD).
\]
In the case under consideration we obtain a similar result. Let us denote by
$M^{\ast}$ the complex conjugate operator of $M$:%
\[
M^{\ast}=\beta\sqrt{\varepsilon\mu}\partial_{t}D+\sqrt{\varepsilon\mu}%
\partial_{t}+iD.
\]
For simplicity we consider now a sourceless situation. In this case the
equations (\ref{wave1}) and (\ref{wave2}) are homogeneous and can be
represented as follows
\[
MM^{\ast}\overrightarrow{U}(t,x)=0,
\]
where $\overrightarrow{U}\ $stands for $\overrightarrow{E}$ or for
$\overrightarrow{H}$.

\section{Fundamental solution of the operator $M$}

We will construct a fundamental solution of the operator $M$ using the results
of the previous section and well known facts from quaternionic analysis.
Consider the equation%
\[
(\beta\sqrt{\varepsilon\mu}\partial_{t}D+\sqrt{\varepsilon\mu}\partial
_{t}-iD)f(t,x)=\delta(t,x).
\]
Applying the Fourier transform $\mathcal{F}$ with respect to the time-variable
$t$ we obtain%
\[
(\beta\sqrt{\varepsilon\mu}i\omega D+\sqrt{\varepsilon\mu}i\omega
-iD)F(\omega,x)=\delta(x),
\]
where $F(\omega,x)=\mathcal{F}\{f(t,x)\}=\int_{-\infty}^{\infty}%
f(t,x)e^{-i\omega t}dt.$ The last equation can be rewritten as follows%
\[
(D+\alpha)(\beta\sqrt{\varepsilon\mu}\omega-1)iF(\omega,x)=\delta(x),
\]
where $\alpha=\frac{\sqrt{\varepsilon\mu}\omega}{\beta\sqrt{\varepsilon\mu
}\omega-1}.$ The fundamental solution of $D_{\alpha}$ is given by
(\ref{fund_d}), so we have%
\[
(\beta\sqrt{\varepsilon\mu}\omega-1)iF(\omega,x)=(\alpha+\frac{x}{\left|
x\right|  ^{2}}-i\alpha\frac{x}{\left|  x\right|  })\Theta_{\alpha}(x),
\]
from where%

\[
F(\omega,x)=\left[  \frac{i\sqrt{\varepsilon\mu}\omega}{(\beta\sqrt
{\varepsilon\mu}\omega-1)^{2}}\left(  1-\frac{ix}{\left|  x\right|  }\right)
+\frac{ix}{\left|  x\right|  ^{2}}\frac{1}{\beta\sqrt{\varepsilon\mu}\omega
-1}\right]  \frac{e^{i\left|  x\right|  \frac{\sqrt{\varepsilon\mu}\omega
}{\beta\sqrt{\varepsilon\mu}\omega-1}}}{4\pi\left|  x\right|  }.
\]
We write it in a more convenient form
\[
F(\omega,x)=\left(  \frac{1}{\left(  \omega-a\right)  ^{2}}A\left(  x\right)
+\frac{1}{\omega-a}B\left(  x\right)  \right)  E\left(  x\right)
e^{\frac{ic(x)}{\omega-a}},
\]
where $a=\frac{1}{\beta\sqrt{\varepsilon\mu}}$, $c\left(  x\right)
=\frac{\left|  x\right|  }{\beta^{2}\sqrt{\varepsilon\mu}}$, $E\left(
x\right)  =\frac{e^{\frac{i\left|  x\right|  }{\beta}}}{4\pi\left|  x\right|
},$%
\[
A\left(  x\right)  =\frac{i}{\beta^{3}\varepsilon\mu}\left(  1-\frac
{ix}{\left|  x\right|  }\right)  ,\qquad B\left(  x\right)  =\frac{i}%
{\beta\sqrt{\varepsilon\mu}}\left(  \frac{1}{\beta}\left(  1-\frac{ix}{\left|
x\right|  }\right)  +\frac{x}{\left|  x\right|  ^{2}}\right)  .
\]
In order to obtain the fundamental solution $f(t,x)$ we should apply the
inverse Fourier transform to $F(\omega,x)$. Among different regularizations of
the resulting integral we should choose the one leading to a fundamental
solution satisfying the causality principle, that is vanishing for $t<0$. Such
an election is done by introducing of a small parameter $y>0$ in the following way%

\begin{equation}
f(t,x)=\lim_{y\rightarrow0}\mathcal{F}^{-1}\left\{  F(z,x)\right\}  \label{f}%
\end{equation}
where $z=\omega-iy$. This regularization is in agreement with the condition
$\operatorname*{Im}\alpha\geq0$. We have%
\begin{equation}
\mathcal{F}^{-1}\left\{  F(z,x)\right\}  =\frac{1}{2\pi}\int_{-\infty}%
^{\infty}\left(  \frac{1}{\left(  \omega-a_{y}\right)  ^{2}}A\left(  x\right)
+\frac{1}{\omega-a_{y}}B\left(  x\right)  \right)  E\left(  x\right)
e^{\frac{ic(x)}{\omega-a_{y}}}e^{i\omega t}d\omega\label{fund}%
\end{equation}
where $a_{y}=a+iy$. Expression (\ref{fund}) includes two integrals of the form%

\[
I_{k}=\frac{1}{2\pi}\int_{-\infty}^{\infty}\frac{e^{\frac{ic}{\omega-a_{y}}%
}e^{i\omega t}}{\left(  \omega-a_{y}\right)  ^{k}}d\omega,\quad k=1,2
\]
where $c=c(x)$. We have
\begin{equation}
I_{k}=\frac{1}{2\pi}\sum_{j=0}^{\infty}\left(  \frac{\left(  ic\right)  ^{j}%
}{j!}\int_{-\infty}^{\infty}\frac{e^{i\omega t}d\omega}{\left(  \omega
-a_{y}\right)  ^{j+k}}\right)  . \label{Ik}%
\end{equation}
Denote
\[
I_{k,j}(t)=\int_{-\infty}^{\infty}\frac{e^{i\omega t}d\omega}{\left(
\omega-a_{y}\right)  ^{j+k}}.
\]
For $k=1$ and $j=0$ we obtain (see, e.g., \cite[Sect. 8.7]{Brem})
\[
I_{1,0}(t)=2\pi iH(t)e^{ita_{y}}%
\]
where $H$ is the Heaviside function. For all other cases, that is for $k=1$
and $j=\overline{1,\infty}$ and for $k=2$ and $j=\overline{0,\infty}$ \ we
have that $j+k\geq2$ and the integrand in (\ref{Ik}) has a pole at the point
$a_{y}$ of order $j+k$. Using a result from the residue theory \cite[Sect.
4.3]{Derrick} we obtain%
\[
I_{k,j}(t)=2\pi i\operatorname*{Res}_{a_{y}}\frac{e^{i\omega t}}{\left(
\omega-a_{y}\right)  ^{j+k}}\quad\text{for }t\geq0\text{ and }j+k\geq2.
\]
Consider%
\[
\operatorname*{Res}_{a_{y}}\frac{e^{i\omega t}}{\left(  \omega-a_{y}\right)
^{j+k}}=\frac{1}{(j+k-1)!}\lim_{\omega\rightarrow a_{y}}\frac{\partial
^{j+k-1}}{\partial\omega^{j+k-1}}e^{i\omega t}=\frac{(it)^{j+k-1}e^{ia_{y}t}%
}{(j+k-1)!}\quad\text{for }t\geq0
\]
and $j+k\geq2.$

For $t<0$ we have that $I_{k,j}(t)$ is equal to the sum of residues with
respect to singularities in the lower half-plane $y<0$ which is zero because
the integrand is analytic there. Thus we obtain%
\[
I_{k,j}(t)=2\pi iH(t)\frac{(it)^{j+k-1}}{(j+k-1)!}e^{ia_{y}t}.
\]
Substitution of this result into (\ref{Ik}) gives us%
\[
I_{1}=iH(t)e^{ia_{y}t}\sum_{j=0}^{\infty}\frac{(-ct)^{j}}{j!j!}\qquad
\text{and}\qquad I_{2}=-H(t)e^{ia_{y}t}t\sum_{j=0}^{\infty}\frac{(-ct)^{j}%
}{j!(j+1)!}.
\]
Now using the series representations of the Bessel functions $J_{0}$ and
$J_{1}$ (see e.g. \cite[Chapter 5]{Vladimirov}) we obtain
\[
I_{1}=iH(t)e^{ia_{y}t}J_{0}\left(  2\sqrt{ct}\right)  \text{\quad and\quad
}I_{2}=-H(t)\sqrt{\frac{t}{c}}e^{ia_{\epsilon}t}J_{1}\left(  2\sqrt
{ct}\right)  .
\]
Substituting these expressions in (\ref{fund}) and then in (\ref{f}) we arrive
at the following expression for $f$:%

\[
f(t,x)=H(t)e^{iat}E\left(  x\right)  \left(  -A\left(  x\right)  \sqrt
{\frac{t}{c}}J_{1}\left(  2\sqrt{ct}\right)  +iB\left(  x\right)  J_{0}\left(
2\sqrt{ct}\right)  \right)  .
\]
Finally we rewrite the obtained fundamental solution of the operator $M$ in
explicit form:%
\begin{align*}
f(t,x)  &  =H(t)\frac{e^{\frac{it}{\beta\sqrt{\varepsilon\mu}}}}{\beta
\sqrt{\varepsilon\mu}}\left(  \mathcal{K}_{\frac{1}{\beta}}(x)J_{0}\left(
\frac{2\sqrt{t\left|  x\right|  }}{\beta\left(  \varepsilon\mu\right)
^{\frac{1}{4}}}\right)  \right. \\
&  \left.  +\frac{i\Theta_{\frac{1}{\beta}}(x)}{\beta\left(  \varepsilon
\mu\right)  ^{\frac{1}{4}}}\left(  1-\frac{ix}{\left|  x\right|  }\right)
\sqrt{\frac{t}{\left|  x\right|  }}J_{1}\left(  \frac{2\sqrt{t\left|
x\right|  }}{\beta\left(  \varepsilon\mu\right)  ^{\frac{1}{4}}}\right)
\right)  .
\end{align*}

Let us notice that $f$ fulfills the causality principle requirement which
guarantees that its convolution with the function from the right-hand side of
(\ref{Amax}) gives us the unique physically meaningful solution of the
inhomogeneous Maxwell system (\ref{Max1})-(\ref{Max3}) in a whole space.

\begin{acknowledgement}
The authors wish to express their gratitute to CONACYT for the support of this
work via the grant C\'{a}tedra Patrimonial No. 010286 and a Research Project.
\end{acknowledgement}


\bigskip

\begin{thebibliography}{99}                                                                                               %


\bibitem {AMS}Athanasiadis C, Martin P and Stratis I \emph{Electromagnetic
scattering by a homogeneous chiral obstacle: boundary integral equations and
low-chirality approximations. }SIAM J. Appl. Math. \textbf{59 (}1999), 1745-1762.

\bibitem {ARS}Athanasiadis C, Roach G and Stratis I \emph{A time domain
analysis of wave motions in chiral materials.} Math. Nachr. \textbf{250}
(2003), 3-16.

\bibitem {Brem}Bremermann H \emph{Distributions, Complex Variables, and
Fourier Transforms. }Addison-Wesley Publ., 1965.

\bibitem {Derrick}Derrick W \emph{Complex Analysis and Applications.
}Wadsworth, Inc., 1984.

\bibitem {GS2}G\"{u}rlebeck K and Spr\"{o}{ss}ig W \emph{Quaternionic and
Clifford Calculus for Physicists and Engineers.} John Wiley \& Sons, 1997.

\bibitem {Imaeda}Imaeda K \emph{A new formulation of classical
electrodynamics.} Nuovo Cimento B \textbf{32} (1976), 138-162.

\bibitem {KKR}Khmelnytskaya K V, Kravchenko V V and Rabinovich V S
\emph{M\'{e}todos cuaterni\'{o}nicos para los problemas de propagaci\'{o}n de
ondas electromagn\'{e}ticas producidas por fuentes en movimiento.
}Cient\'{\i}fica: The Mexican Journal of Electromechanical Engineering
\textbf{5} (2001), 143-146.

\bibitem {KKR1}Khmelnytskaya K V, Kravchenko V V and Rabinovich V S
\emph{Quaternionic fundamental solutions for electromagnetic scattering
problems and application.} Zeitschrift f\"{u}r Analysis und ihre Anwendungen
\textbf{22} (2003), 147--166. \ 

\bibitem {Krdep}Kravchenko V~V \emph{On the relation between holomorphic
biquaternionic functions and time-harmonic electromagnetic fields.} Deposited
in UkrINTEI, $29.12.1992,\#2073-Uk-92,18$ pp. (Russian).

\bibitem {AQA}Kravchenko V~V \emph{Applied quaternionic analysis. }
Heldermann-Verlag, Research and Exposition in Mathematics Series, v. 28, 2003.

\bibitem {Lak}Lakhtakia A \emph{Beltrami fields in chiral media.} World
Scientific, 1994.

\bibitem {LSihvola}Lindell I V, Sihvola A H, Tretyakov S A and Viitanen A J
\emph{Electromagnetic waves in chiral and bi-isotropic media.} Artech House, 1994.

\bibitem {Shn}Shneerson M S \emph{Maxwell's equations and functional-invariant
solutions of the wave equation.} Differencialnye Uravneniya \textbf{4} (1968),
743-758 (Russian).

\bibitem {Vladimirov}Vladimirov V S \emph{Equations of mathematical physics.}
Nauka, 1984 (Russian); Engl. transl. of the first edition: Marcel Dekker, 1971.
\end{thebibliography}
\end{document}